\documentclass[twocolumn,english,twocolumn,showpacs,preprintnumbers,amsmath,amssymb,superscriptaddress]{revtex4}
\usepackage[T1]{fontenc}
\usepackage[latin1]{inputenc}
\usepackage{graphicx}

\makeatletter

\makeatletter
\makeatletter
\makeatletter

\makeatletter

\makeatletter

\makeatletter

\makeatletter

\makeatletter

\@ifundefined{definecolor}
 {\@ifundefined{definecolor}
 {\@ifundefined{definecolor}
 {\usepackage{color}}{}
}{}
}{}

\usepackage{dcolumn}

\usepackage{bm}

\newcommand{\pt}{ p_{\rm t}}
\newcommand{\ie}{{\it i.e.}}

\def\lsim{\mathrel{\rlap{\lower4pt\hbox{\hskip1pt$\sim$}}
    \raise1pt\hbox{$<$}}}         %less than or approx. symbol
\def\gsim{\mathrel{\rlap{\lower4pt\hbox{\hskip1pt$\sim$}}
    \raise1pt\hbox{$>$}}}         %greater than or approx. symbol

\makeatother

\makeatother

\makeatother

\makeatother

\def\citet{\cite}

\makeatother

\makeatother

\usepackage{babel}

\begin{document}

\title{Direct photons at low transverse momentum -- a QGP signal in pp collisions
at LHC}

\author{Fu-Ming Liu}

\affiliation{Institute of Particle Physics \rm and Key laboratory of Quark \& Lepton Physics (Ministry of Education), Central China Normal University, Wuhan,
China}

\author{Klaus Werner}

\affiliation{Laboratoire SUBATECH, University of Nantes - IN2P3/CNRS - Ecole desMines,
Nantes, France}

\date{\today}

\begin{abstract}
We investigate photon production in a scenario of quark-gluon plasma
formation in proton-proton scattering at 7~TeV.
It is shown that thermal photon yields increase quadratically with
the charged particle multiplicity. This gives an enhanced weight to
high multiplicity events, and leads to an important photon production
even in minimum bias events, where the thermal photons largely dominate
over the prompt ones at transverse momentum values smaller than 10
GeV/c.
\end{abstract}
\maketitle

%\section{Introduction}

The QCD phase diagram tells us that a new kind of matter, the quark
gluon plasma (QGP), can be formed at very high temperature or at very
high density via high energy collisions. Such matter has probably
been observed in heavy ion collisions at the Relativistic Heavy Ion Collider (RHIC). At least there
seems to be no doubt that matter expands collectively, governed by
a hydrodynamical expansion. This has essentially been proved based
on studies of azimuthal anisotropies, where for example elliptical
flow and in particular its mass dependence can hardly be explained
without considering strong collectivity. It was for the first time
in heavy ion physics that hydrodynamic models were able to describe
such nontrivial features correctly, and it seems also clear that the
corresponding liquid is almost perfect, in the sense of having very
small viscosity.  

Originally hydrodynamics was only thought to present a valid description
for almost central collisions of heavy nuclei, where the volume is
(relatively) big. But it seems that this approach works very well
for all centralities. There is also no fundamental difference observed
between CuCu and AuAu, although the copper system is much smaller.
Systems much smaller than central AuAu also fit well into
this fluid picture. Finally it is more and more accepted that the
famous ridge structure observed in angle-rapidity dihadron correlation
\citet{cms_ridge} is due to fluctuating initial conditions, which
are subsequently transformed into collective flow \citet{epos2}.
Here, the relevant scale for applying hydrodynamics is not the nuclear
size, but the size of the fluctuations, which is typically 1-2 fermis.

Is QGP formation a nuclear phenomenon? Or can it be formed in pp scattering, 
as proposed originally in refs. \cite{oldhydro1, Belenkij:1956cd, oldhydro2} and advocated more recently in
 \citet{oldhydro3,oldhydro4,epos2,ridge}?
Based on the above discussion, there is no reason not to treat proton-proton
scattering in the same way as heavy ions, namely incorporating a hydrodynamical
evolution. This approach makes clear predictions for many variables,
so the Nature will tell us whether the approach is justified or not.
Therefore it will be extremely interesting to think about the implications
of such a mini QGP, how such a small system can equilibrate so quickly,
and so on. It would be an enormous waste of opportunities, not to
consider this possibility, since a vast amount of proton-proton data
will be available very soon, concerning all kinds of observables.

What makes pp scattering at LHC energies interesting in this respect,
is the fact that at this high energy multiple scattering becomes very
important, where a large number of scatterings amounts to a large
multiplicity. In such cases, very large energy densities occur, even
bigger than the values obtained in heavy ion collisions at RHIC --
but in a smaller volume. Several authors discussed already the possibility
of a hydrodynamical phase in pp collisions at the LHC, to explain
the ridge correlation \citet{Khachatryan:2010gv,bozek,epos2}, or
to predict elliptic flow \citet{solana,david,prasad,kodama}.

In heavy ion physics, one of the possible {}``signals'' of QGP formation is photon
production, since a hot plasma radiates a large amount of {}``thermal''
photons, which dominate the spectra at small transverse momenta, whereas
large momenta are dominated by photons from hard processes in nucleon-nucleon
scatterings. Such a {}``low pt enhancement'' has been observed at
RHIC, see \citet{PHENIX10,Liu:2008eh}. Photons are an interesting
plasma signal, since they are emitted from the interior of the hot
matter, and do not interact any more, contrary to hadronic observables.
The only question here: are there kinematic windows (pt range) where thermal
photons can clearly be distinguished from other sources? 

In this paper we will study the possibility of a QGP formation in
proton-proton scattering, by making use of the properties of photons
as a signature. So we will discuss direct photon production, both prompt
photons and thermal photons in details, in pp collisions at 7~TeV. 

\begin{figure}
\includegraphics[scale=0.85]{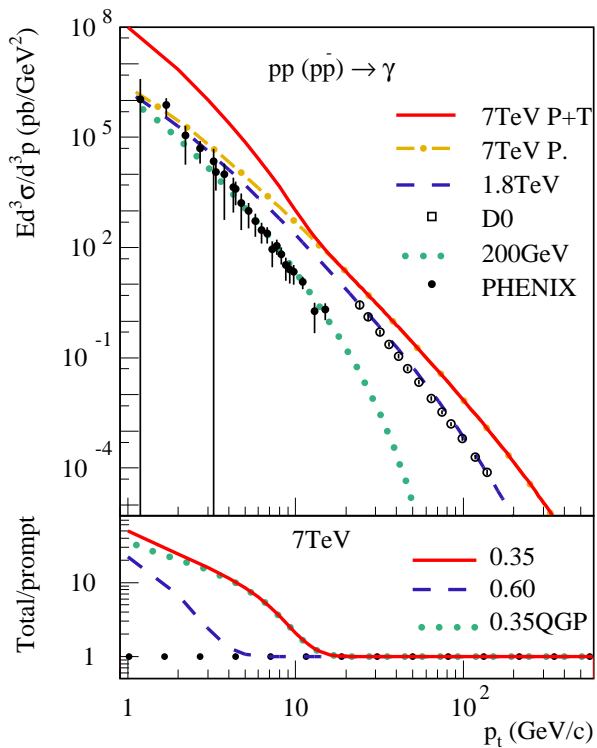}

\caption{\label{fig:Fig1} (Color Online) Upper panel: Prompt photons (dashed-dotted
line) and the sum of prompt photons and thermal photons (full line)
in $pp$ at 7~TeV. We also show prompt photon production in $pp$
at 200GeV (dotted line) and in $p\bar{p}$ at 1.8~TeV (dashed line)
compared to PHENIX data \citet{PHENIX07PRL} (full cycles) and D0
data \citet{Abachi:1996qz} (open circles). Lower panel: the ratio
of the total photon production (prompt + thermal) to the prompt photons
for $pp$ collisions at 7~TeV. See text. }

\end{figure}

But before discussing the details, we want to show in Fig.~\ref{fig:Fig1}
the final result. We show first of all prompt photon calculations
at 200~GeV and at 1.8~TeV, compared to data from RHIC and Tevatron.
There seems to no need for thermal photon production at 200~GeV,
whereas at Tevatron, there are no data at low $\pt$, and therefore
no conclusion can be drawn on the question of thermal photon production.
Interesting are the results we obtain for pp scattering at 7~TeV:
we show in Fig.~\ref{fig:Fig1} the prompt photons (dashed-dotted
line) and the full contribution, prompt plus thermal ones (full line).
Here we get an important contribution from thermal photons, which
largely dominate the spectrum up to roughly 10~GeV/c. Very important
here is the very early stage, where the temperatures are highest.
To show this, we plot in the lower panel the ratio of the total photon
production (prompt + thermal) to the prompt photons. The full line
corresponds to the full line in the upper panel -- the default calculation corresponding to a hydrodynamical evolution starting at $\tau_{0}$
= 0.35~fm/c. The dashed line is the result of a calculation starting
at $\tau_{0}$ = 0.60~fm/c. The thermal contribution is considerably
reduced, but still important. Since early production is very important,
it is clear that photons from the hadronic phase are negligible. The
dotted line in the lower panel corresponds to the default calculation
again, but counting only photons from the QGP phase. So the main message
is that we get a very important thermal contribution for transverse
momentum up to 10~GeV/c, due to the very early emission from the
QGP phase. But why is the contribution so big, and why do we get such
a big effect only at very high energies (why not already in p+p collisions at RHIC)?
To answer these questions, we have to discuss more details, as will
be done in the following chapters.

%\section{Prompt photons}

The production of high-$\pt$ prompt photons in $pp$($p\bar{p}$)
collisions is an important testing ground for perturbative QCD. Primarily
due to the relatively clean signal provided by photons and their point-like
coupling to quarks, this enables a probe of the dynamics of the underlying
hard scattering subprocesses that involve strong interactions. The
cross section for the fully inclusive production of a single prompt
photon schematically reads \citet{Owens1987} \begin{eqnarray*}
\frac{d\sigma^{{\rm prompt}}}{dyd^{2}\pt} & = & \sum_{{\displaystyle ab}}\int dx_{a}dx_{b}G_{a}(x_{a},M^{2})G_{b}(x_{b},M^{2})\frac{\hat{s}}{\pi}\\
 & \times & \delta(\hat{s}+\hat{t}+\hat{u})[\frac{d\sigma}{d\hat{t}}(ab\rightarrow\gamma+X)\\
 & + & K\sum_{c}\frac{d\sigma}{d\hat{t}}(ab\rightarrow cd)\int dz_{c}\frac{1}{z_{c}^{2}}D_{\gamma/c}(z_{c},Q^{2})]\end{eqnarray*}
 where $G_{a}(x_{a},M^{2})$ is parton distribution functions (PDF)
in proton, the elementary processes $ab\rightarrow\gamma+X$ are Compton
scattering $qg\rightarrow\gamma q$ and annihilation $q\bar{q}\rightarrow g\gamma$
and the second term covers high order contribution with photon fragmentation
functions $D_{\gamma/c}(z_{c},Q^{2})$ being the probability for obtaining
a photon from a parton $c$ which carries a fraction $z$ of the parton's
momentum. In our calculation, MRST2001 \citet{Martin:2001es} PDF
is employed and $K$=2 is used to take into account high order contribution
of hard parton production. The obtained $\pt$ spectrum of prompt
photons at three energies are the curves shown earlier in Fig.~1.

In the following, we will introduce another source of photon production,
completely new in the field of proton-proton scattering, which will
substantially modify the spectra in the low $\pt$ region. Therefore
we have to discuss the question of how well we understand {}``normal''
photon production in this area. There are several factors, like the
theoretical scales, parton distribution functions, and so on. Extensive
investigation on the effects from different renormalization scales
and factorization scales and from different PDF such as GRV-94, MSRT
and CETQ-2M have been done\citet{photon97}. More recently PDF are
discussed in the context of parton saturation: a slightly lower production
in the low $\pt$ region may expected when using corresponding PDF.
Another topic to be discussed concerns higher order contributions,
which get more important with increasing collision energy \citet{Aurenche:2006vj}
and may enhance low $\pt$ photon production. However, the resummation
calculation of \citet{Diana:2010ef} shows only a few percent increase.

%\section{Thermal Photons in a Multiple Scattering Approach}

\begin{figure}
\includegraphics[scale=0.85]{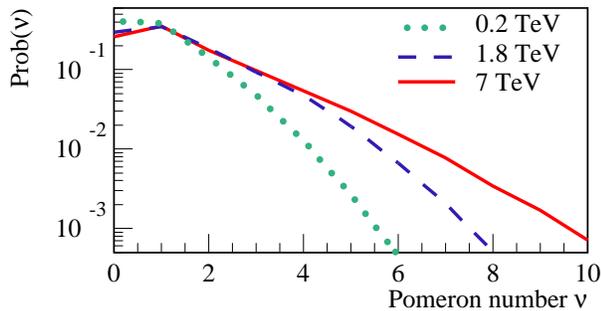}

\caption{\label{fig:nuprob} (Color Online) Probability distribution of Pomeron
number $\nu$ at three collision energies.}

\end{figure}

As discussed earlier, crucial for our discussion is the fact that
at LHC energies multiple scattering in the spirit of the Gribov-Regge
approach becomes important, and therefore the number $\nu$ of Pomerons
is a key quantity for each event. Whereas in heavy ion collisions
the centrality is used to define event classes, we classify here events
according the Pomeron number $\nu$. In the so-called Eikonal approximation,
the probability of $\nu$ Pomerons reads $Prob(\nu)\propto\frac{\chi^{\nu}}{\nu!},$
where Pomerons are treated as identical, and $\chi$ depends on nothing
but the collision energy $\sqrt{s}$. However, this approximation
ignores two facts: the collision energy is shared between the $\nu$
Pomerons, and initial valence quarks of the two colliding protons
destroy the simplified picture of having identical Pomerons. A more
sophisticated multiple scattering theory\citet{Drescher:2000ha,Liu:2003wja}
was developed to improve this approximation, and the resulting Pomeron
distribution is shown in Fig.\ref{fig:nuprob}, for $pp$ collisions
at $\sqrt{s}=$200GeV (dotted line), $p\bar{p}$ at 1.8~TeV (dashed
line) and $pp$ at 7~TeV (solid line).

From the multi-string configurations for a given $\nu$, the energy-momentum
tensor \begin{eqnarray}
T^{\mu\nu}=(e+P)u^{\mu}u^{\nu}-Pg^{\mu\nu}\end{eqnarray}
 at some initial time $\tau_{0}$ is obtained, where 
 $e$ is energy density, $P$ is pressure, $u^{\mu}$ is local four fluid velocity
 and $g^{\mu\nu}=\rm{diag}(1,-1,-1,-1)$ is the metric tensor. 
The subsequent evolution is governed by conservation laws of energy and momentum, 
\begin{eqnarray}\partial_{\mu}T^{\mu\nu}=0. \label{eq:hydro}\end{eqnarray} 
We use an equation-of-state which is compatible with lattice gauge results 
of ref. \citet{lattice}.
The above equations \citet{hydroequation} are solved in full 3D space ($\tau$, $x$, $y$,
$\eta_{s}$) where $\tau$, $\eta_{s}$, $x$, and $y$ are the proper
time, space-time rapidity, the two transverse coordinates, to obtain
energy density $e$, pressure $P$, and local four fluid velocity
$u^{\mu}$, respectively.
 With $\tau_{0}$=0.35fm/c, the obtained
plasma evolution successfully explained\citet{epos2,ridge}
multiplicity distributions, rapidity distributions, $\pt$ spectra
and mean $\pt$ dependence of multiplicity, Bose-Einstein correlations,
and the ridge phenomenon of charged hadrons. Contrary to \citet{epos2,ridge},
we employ here the initial condition for each given Pomeron number
$\nu$.

\begin{figure}
\includegraphics[scale=0.85]{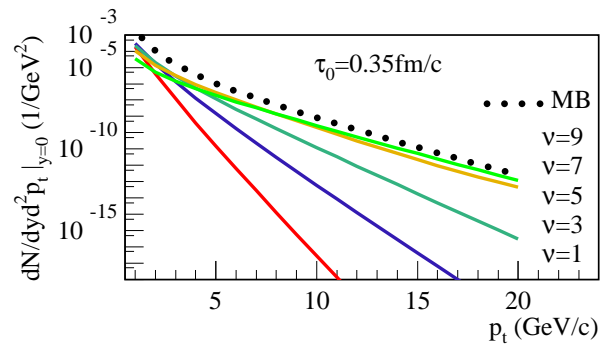}

\caption{\label{fig:Fig2} (Color Online) The thermal spectra given $\nu$
(solid lines) and the mini-bias case (dotted line). The different
solid lines correspond to (from bottom to top) $\nu=1$, 3, 5, 7,
9.}

\end{figure}

Now thermal photon emission can be treated in the same way as in heavy
ion collisions \citet{Liu:2008eh}, $\ie$, the transverse momentum
spectra of thermal photons at a given $\nu$ can be written as \begin{equation}
\frac{dN}{dy\, d^{2}p_{t}}(\nu)=\int d^{4}x\,\Gamma(E^{*},T)\label{eq1}\end{equation}
 with $\Gamma(E^{*},T)$ being the Lorentz invariant thermal photons
emission rate which covers the contributions from the QGP phase \citet{AMY}
and HG phase \citet{MYM}, $d^{4}x=\tau\, d\tau\, dx\, dy\, d\eta_{s}$
being the volume-element, and $E^{*}=p^{\mu}u_{\mu}$ the photon energy
in the local rest frame. Here $p^{\mu}$ is the photon's four momentum
in the laboratory frame, $T$ and $u^{\mu}$ are the temperature and
the local fluid velocity, respectively, obtained from solving eq.(\ref{eq:hydro})
for each Pomeron number $\nu$. The Mini-bias thermal contribution
reads

\[
\frac{dN}{dy\, d^{2}p_{t}}({\rm MB})=\sum_{\mu}\frac{dN}{dy\, d^{2}p_{t}}(\nu)*Prob(\nu).\]
 In Fig.\ref{fig:Fig2} we plot the thermal spectra from given $\nu$,
including the factor $Prob(\nu)$ (solid lines) and the minimum-bias
case (dotted line).

The corresponding thermal cross section $d\sigma^{{\rm thermal}}/dy\, d^{2}p_{t}$
is then simply obtained by multiplying with the inelastic pp cross
section (we use $\sigma_{pp}$=63.2~mb). What is very interesting
is the fact that the $\pt$ region of 5-10 GeV/c is completely dominated
by large values of $\nu$. The smallness of $Prob(\nu)$ for $\nu$
around 7-9 is compensated by the very hard pt spectrum, as compared
to small $\nu$ values, due to the very high energy densities for
large $\nu$. This provides the very fortunate situation that {}``QPG
effects'' which are obviously more developed for large $\nu$, are
already visible in minimum bias spectra. From Figs. 2 and 3 it is
also clear why we do not expect such a big effect at RHIC. Here large
values of $\nu$ are strongly suppressed, having $\nu=9$ for example
is practically impossible, and therefore the large $\nu$ plasma effect
cannot be seen.

\begin{figure}
\includegraphics[scale=0.85]{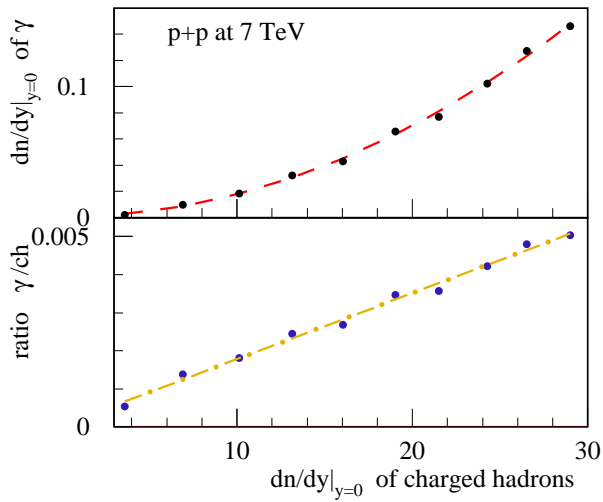}

\caption{\label{fig:Fig4b} (Color Online) Upper panel: the rapidity plateau
height of charged hadrons (squares) and thermal photons (dots) are
plotted versus $\nu$. The yellow bands on the left represent minimum-bias
results. In the lower panel, we plot the ratio of the plateau heights
of photons and charged particles (triangles). The dashed lines are
used to guide the eye. }

\end{figure}

Whereas the Pomeron number $\nu$ is the important quantity for theoretically
defining event classes, we have to finally use real observables. Fortunately
there is a strong relation between Pomeron number and the
charged particle multiplicity. Fitting the values of the pseudorapidity
density $dn/dy(y=0)$ for given $\nu$, we get: $$dn/dy(0)=2.8147 \nu +4.3477.$$  In the
upper panel of Fig.~\ref{fig:Fig4b}, the (pseudo)rapidity density
$dn^{\gamma}/dy(0)$ of thermal photons versus $dn/dy(0)$ is plotted.
Because photons are massless, the result is very sensitive to the
lower limit in the transverse momentum integration\citet{Liu:2009kta},
which is taken to be zero in our calculation. In the lower panel we
plot the ratio $dn^{\gamma}/dy(0)$ to $dn/dy(0)$. The lines in both
panels are used to guide the eye. One can clearly see the linear dependence
in the lower panel, which mean photon production increases quadratically
with the charged particle multiplicity! This is understandable, because
photon emission from a QGP is a volume process, integrated over time
while hadrons are emitted from the freeze-out hypersurface, corresponding
to a narrow window in proper-time. This strong increase of photon
production with multiplicity is another reason why photons are a very
good probe to investigate QGP production in high multiplicity pp events.

%\section{Discussion and Conclusion}

In summary, we illustrate that direct photon production at low and
intermediate $\pt$ is a good signal of QGP formation in pp scattering.
The reason is that photon production is predicted to increase quadratically
with the the charged particle multiplicity. Therefore high multiplicity
events contribute considerably even to minimum bias photon production,
and as a consequence at low and intermediate $\pt$ values, the thermal
production should be by far dominant compared to prompt photons.

%\begin{acknowledgments}
This work is supported by the Natural Science Foundation of China
under the project No.~10975059. 
%\end{acknowledgments}


\begin{thebibliography}{10}
\bibitem{cms_ridge}CMS Collaboration, JHEP 1009:091,2010

\bibitem{epos2}K.$\,$Werner, Iu.$\,$Karpenko, T.$\,$Pierog, M.
Bleicher, K. Mikhailov, Phys. Rev. C 82, 044904 (2010)

\bibitem{oldhydro1}L. D. Landau, Izv. Akad. Nauk SSSR 17, 51 (1953).

 \bibitem{Belenkij:1956cd}
  S.~Z.~Belenkij and L.~D.~Landau,
  %``Hydrodynamic theory of multiple production of particles,''
  Nuovo Cim.\ Suppl.\  {\bf 3S10}, 15 (1956)
 [Usp.\ Fiz.\ Nauk {\bf 56}, 309 (1955)].
  %%CITATION = UFNAA,56,309;%%

\bibitem{oldhydro2}P. Carruthers, Annals N.Y. Acad. Sci. 229, 91
(1974).

\bibitem{oldhydro3}M. Luzum, P. Romatschke, Phys. Rev. Lett. 103,
262302 (2009)

\bibitem{oldhydro4}S. Prasad et. al. arXiv: 0910.4844, G. Orta et
al. arXiv: 0911.2392

\bibitem{ridge}K. Werner, Iu. Karpenko, and T. Pierog, Phys. Rev.
Lett. 106, 122004 (2011)]


\bibitem{Khachatryan:2010gv} V.~Khachatryan \textit{et al.} {[}CMS
Collaboration], %``Observation of Long-Range Near-Side Angular Correlations in Proton-Proton
 %Collisions at the LHC,''
 eConf \textbf{C990809}, 3 (2000) {[}arXiv:1009.4122 {[}hep-ex]].
%%CITATION = JHEPA,1009,091;%%


\bibitem{bozek}P. Bozek, arXiv:1010.0405.

\bibitem{solana}J. Casalderrey-Solana, U. A. Wiedemann, Phys.Rev.Lett.104:
102301(2010).

\bibitem{david}D. d'Enterria, G.Kh. Eyyubova, V.L. Korotkikh, I.P.
Lokhtin, S.V. Petrushanko, L.I. Sarycheva, A.M. Snigirev, Eur. Phys.
J. C 66, 173 (2010).

\bibitem{prasad}S. K. Prasad, Victor Roy, S. Chattopadhyay, A. K.
Chaudhuri, Phys.Rev.C82:024909,2010.

\bibitem{kodama}G. Ortona, G. S. Denicol, Ph. Mota, T. Kodama, arXiv:0911.5158.


\bibitem{PHENIX10} A.~Adare \textit{et al.} {[}PHENIX Collaboration],
%``Enhanced production of direct photons in Au+Au collisions at sqrt(s_NN)=200
 %GeV and implications for the initial temperature,''
 Phys.\ Rev.\ Lett.\ \textbf{104}, 132301 (2010), %%CITATION = ARXIV:0804.4168;%% \\



\bibitem{Liu:2008eh} F.~M.~Liu, T.~Hirano, K.~Werner and Y.~Zhu,
%``Centrality-dependent direct photon pt spectra in Au+Au collisions at
 %RHIC,''
 Phys.\ Rev.\ C \textbf{79} (2009) 014905. F.~M.~Liu, %``A Theoretic Review of Centrality-Dependent Direct Photon pt spectra in
 %Au+Au Collisions at 200 GeV,''
 arXiv:1012.0086 {[}nucl-th]. %%CITATION = ARXIV:1012.0086;%%



\bibitem{PHENIX07PRL} S.~S.~Adler \textit{et al.} {[}PHENIX Collaboration],
Phys. Rev. Lett. \textbf{98}, 012002 (2007).

\bibitem{Abachi:1996qz} S.~Abachi \textit{et al.} {[}D0 Collaboration],
%``Isolated photon cross-section in the central and forward rapidity regions
 %in $p\bar{p}$ collisions at $\sqrt{s} = 1.8$ TeV,''
 Phys.\ Rev.\ Lett.\ \textbf{77}, 5011 (1996) {[}arXiv:hep-ex/9603006].
%%CITATION = PRLTA,77,5011;%%

\bibitem{Owens1987}J.F.~Owens, Rev. Mod. Phys. 59, 465 (1987).

\bibitem{Martin:2001es} A.~D.~Martin, R.~G.~Roberts, W.~J.~Stirling
and R.~S.~Thorne, %``MRST2001: Partons and $\alpha_s$ from precise deep inelastic scattering and
 %Tevatron jet data,''
 Eur.\ Phys.\ J.\ C \textbf{23}, 73 (2002) {[}arXiv:hep-ph/0110215].
%%CITATION = EPHJA,C23,73;%%

\bibitem{photon97} W.~Vogelsang and M.~R.~Whalley, %``A Compilation of data on single and double prompt photon production in
 %hadron hadron interactions,''
 J.\ Phys.\ G \textbf{23}, A1 (1997). %%CITATION = JPHGB,G23,A1;%%


\bibitem{Aurenche:2006vj} P.~Aurenche, M.~Fontannaz, J.~P.~Guillet,
E.~Pilon and M.~Werlen, %``A New critical study of photon production in hadronic collisions,''
 Phys.\ Rev.\ D \textbf{73}, 094007 (2006) {[}arXiv:hep-ph/0602133].
%%CITATION = PHRVA,D73,094007;%%

\bibitem{Diana:2010ef} G.~Diana, J.~Rojo and R.~D.~Ball, %``High energy resummation of direct photon production at hadronic
 %colliders,''
 Phys.\ Lett.\ B \textbf{693}, 430 (2010) {[}arXiv:1006.4250 {[}hep-ph]].
%%CITATION = PHLTA,B693,430;%%


\bibitem{Drescher:2000ha} H.~J.~Drescher, M.~Hladik, S.~Ostapchenko,
T.~Pierog and K.~Werner, %``Parton-based Gribov-Regge theory,''
 Phys.\ Rept.\ \textbf{350}, 93 (2001) {[}arXiv:hep-ph/0007198].
%%CITATION = PRPLC,350,93;


\bibitem{Liu:2003wja} F.~M.~Liu, J.~Aichelin, M.~Bleicher, H.~J.~Drescher,
S.~Ostapchenko, T.~Pierog and K.~Werner, %``Constraints on models for proton-proton scattering from multistrange baryon
 %data,''
 Phys.\ Rev.\ D \textbf{67}, 034011 (2003). %%CITATION = PHRVA,D67,034011.%%


\bibitem{lattice}S. Borsanyi et al., arXiv:1007.2580.

\bibitem{hydroequation}
Though the viscosity is small as discussed in the introduction, 
one should nevertheless apply viscous hydrodynamics, 
as discussed in ref.\ \cite{viscous} to do controlled calculations.
 
\bibitem{viscous}
B. Schenke, S. Jeon, C. Gale, Phys.Rev.Lett. 106 (2011) 042301.
 

\bibitem{AMY} P.\ Arnold, G.\ Moore and L.G.\ Yaffe, J.\ High
Energy Phys. \textbf{0111}, 057 (2001); J.\ High Energy Phys. \textbf{0112},
9 (2001).

\bibitem{MYM} S.\ Turbide, R.\ Rapp and C.\ Gale, Phys.\ Rev.\ C
\textbf{69}, 014903 (2004).




\bibitem{Liu:2009kta} F.~M.~Liu, T.~Hirano, K.~Werner and Y.~Zhu,
%``Elliptic flow of thermal photons in Au+Au collisions at $\sqrt{s_{NN}}=200$
 %GeV,''
 Phys.\ Rev.\ C \textbf{80}, 034905 (2009) {[}arXiv:0902.1303 {[}hep-ph]].
%%CITATION = PHRVA,C80,034905.%%

\end{thebibliography}
\end{document}